\documentclass[preprintnumbers,amsmath,amssymb,pra]{revtex4}
\usepackage{graphicx}
\usepackage{dcolumn}
\usepackage{bm}
\usepackage{times}
\usepackage[colorlinks,citecolor=blue,linkcolor=red]{hyperref}
\usepackage{color,soul}
\usepackage{comment}
\usepackage{ulem}

\begin{document}

\title{Thermal rectification and heat amplification in a nonequilibrium V-type three-level system}

\author{Chen Wang$^{1,}$}\email{wangchenyifang@gmail.com}
\author{Dazhi Xu$^{2}$}\email{dzxu@bit.edu.cn}
\author{Huan Liu$^{1}$}
\author{Xianlong Gao$^{1}$}
\address{
$^{1}$Department of Physics, Zhejiang Normal University, Jinhua 321004, Zhejiang , P. R. China\\
$^{2}$Department of Physics and Center for Quantum Technology Research,
Beijing Institute of Technology, 5 South Zhongguancun Street, Beijing 100081, China
}

\date{\today}

\begin{abstract}
Thermal rectification and heat amplification are investigated in a nonequilibrium V-type three-level system with quantum interference.
By applying the Redfield master equation combined with full counting statistics, we analyze the steady state heat transport.
The noise-induced interference is found to be able to rectify the heat current,
which paves a new way to design quantum thermal rectifier.
Within the three-reservoir setup, the heat amplification is clearly identified far-from equilibrium,
which is in absence of the negative differential thermal conductance.
\end{abstract}

\maketitle




\section{Introduction}
How to smartly control energy flow and efficiently manipulate logical gates is a challenging problem, ranging from molecular electronics~\cite{maratner2013nn}, spintronics~\cite{izutic2004rmp}, quantum information and computation~\cite{nelsen2011book}.
The electronic diode and electronic transistor, as two main ingredients, have spurred the emergence of semiconductor industry~\cite{dneaman2003book}.
Inspired by these concepts in electronic systems, thermal rectifier (thermal diode) and thermal transistor
have been proposed in phononics~\cite{bli2006apl,nbli2010rmp,jren2015aipadv}.
They constitute the basis of functional thermal devices, realized in quantum dots~\cite{rscheibner2008njp,rsanchez2017prb,grossello2017prb,rsanchez2017njp,amvicioso2018prb,yczhang2018epl}, nanotubes~\cite{cchang2006science}, phase change materials~\cite{pjzwol2012prl},
thermal metamaterials~\cite{tchan2014am,tchan2014prl},
and hybrid normal metal-superconductor nanojunctions~\cite{fgiazotto2013apl,fgiazotto2013apl2,fgiazotto2014apl,fgiazotto2015nn,fgiazotto2016prb}.

Thermal rectification, one of the most fundamental phononic components, is described as a device exhibiting a larger heat flow in one direction than its counterpart in the opposite direction, driven by the thermal gradient.
It is defined as~\cite{truokola2009prb,lfzhang2010prb}
\begin{eqnarray}
\textrm{R}_J=(J_+-J_-)/\max\{J_+,J_-\},
\end{eqnarray}
where $\textrm{R}_J$ denotes the rectification of the current and $J_{{\pm}}$ are heat currents in the forward and backward gradient configurations.
The thermal rectification effect has been intensively investigated in two-terminal phononic lattice~\cite{nbli2010rmp,bwli2004prl,bli2006apl}, spin systems~\cite{lfzhang2009prb,jomiranda2017pre}
and nonequilibrium spin-boson model~\cite{dsegal2005prl}.
It was later extended to three-terminal phononic thermoelectric system~\cite{jhjiang2015prb,jhjiang2016crp} and atomic junctions~\cite{lfzhang2010prb}.
Typically, the quantum rectification can be realized in the asymmetric structures
of quantum systems~\cite{bwli2004prl,bli2006apl,lfzhang2009prb,jomiranda2017pre,lfzhang2010prb,twerlang2014pre,nbli2014sr,zxman2016pre}, different system-bath couplings~\cite{dsegal2005prl,lawu2009prl,lawu2009pre,jren2013prb,jren2013prb2},
or including an additional phonon bath~\cite{jhjiang2015prb} or a probe~\cite{rsanchez2015njp,rsanchez2015pe}.
Recently, the noise-induced interference was unraveled to enhance the quantum coherence
for both transient dynamics~\cite{dps2009pra,pnalbach2010njp,tvtscherbul2014prl,adodin2016jcp,skoyu2017arxiv} and steady state behavior~\cite{swli2015aop,jjun2018scichina,zhwang2018njp,vvkozlov2006pra} in the quantum V-type system.
It was also considered as a novel source to  significantly improve the energy power and efficiency~\cite{moscully,gpanni2010pnas,moscully2011pnas,kedorfman2013pnas,kedorfman2018pre}.
By tieing two seemingly unrelated effects together, i.e., thermal rectification and noise-induced interference, we ask the first question:
{\itshape{will quantum interference exhibit the rectification in the nonequilibrium V-type system}}?

Heat amplification, that a slight change in the base heat current will dramatically change heat currents at the collector and emitter, realizes the thermal transistor~\cite{nbli2010rmp}.
The amplification factor is defined by the ratio
\begin{eqnarray}
\beta_u=|{\partial}J_u/{\partial}J_b|,~u=c,e
\end{eqnarray}
with $J_b$ the base current, $J_c$ the collector current and $J_e$ the emitter current~\cite{bli2006apl}.
Usually, the heat transistor is announced to work as $\beta_{u}>1$.
In previous works, it was widely believed that the negative differential thermal conductance (NDTC) is a compulsory ingredient
of the heat amplification~\cite{pbabd2014prl,kjoulain2015apl,kjoulain2016prl,gtcraven2017prl,cwang2018pra}.
The NDTC is traditionally described by the phenomenon that the heat current decrease by increasing the temperature bias between two baths~\cite{dhhe2009prb,dhhe2010pre,dhhe2014pre,shsu2018arxiv}.
However, in a recent study of phononic thermoelectric system, J. H. Jiang {\itshape{et al}.} proposed that heat amplification can work in linear response regime, even without the NDTC~\cite{jhjiang2015prb}.
Hence, we raise the second question:
{\itshape{based on the V-type system, can we realize the heat amplification far-from equilibrium in absence of the NDTC?}}

To answer these questions, we investigate the steady state heat transfer in a nonequilibrium V-type system,
with the model detailed in Sec.~II~A.
We apply the Redfield scheme to obtain the quantum master equation by including the noise-induced interference, detailed in Sec.~II~B.
The effect of quantum interference on  nonequilibrium steady state coherence is analytically analyzed in Sec.~II~C.
In Sec.~III, we combine the Redfield master equation with full counting statistics~\cite{mesposito2009rmp} to obtain the expression of heat currents.
In Sec.~IV, we study the influence of quantum interference on the thermal rectification within the two-reservoir setup.
In Sec.~V, we investigate the heat amplification in the three-reservoir nonequilibrium V-type system.
Finally, we give a brief summary in Sec.~VI.

\begin{figure}[tbp]
\vspace{-3.5cm}
\includegraphics[scale=0.4]{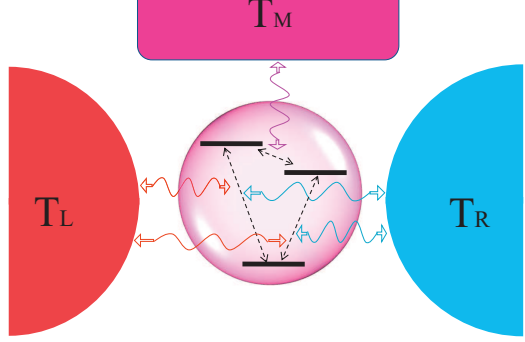}
\caption{(Color online) Schematic diagram of the nonequilibrium V-type three-level system represented by three black horizontal lines,
and the transitions between states are shown as double-arrowed dashed lines.
The red left and blue right half circles are thermal baths, with temperatures $T_L$ and $T_R$, respectively.
The purple upper square is the middle thermal bath with the temperature $T_M$.
The interactions between thermal baths and V-type system are described as the double-arrowed wave lines.
}~\label{fig1}
\end{figure}


\section{Model and Method}
We first describe a V-type system, which interacts with three thermal reservoirs.
Then, we include the Redfield scheme to obtain the dynamical equation of the nonequilibrium V-type system in weak system-bath coupling regime.
Finally, we analyze the effect of the noise-induced interference on the quantum steady state coherence.

\subsection{Nonequilibrium V-type system}
The model to exhibit nonequilibrium  heat transfer through a V-type three-level system interacting with thermal baths in Fig.~\ref{fig1}, is expressed as
$\hat{H}=\hat{H}_s+\hat{H}_b+\sum_{u=L,M,R}\hat{V}_u$.
The three-level system is described as
\begin{eqnarray}~\label{hs0}
\hat{H}_s=\sum_{i=1,2}\varepsilon_i|e_i{\rangle}{\langle}e_i|+\varepsilon_g|g{\rangle}{\langle}g|,
\end{eqnarray}
where $\varepsilon_{1}$ and $\varepsilon_{2}$ are energy levels of two excited states $|e_1{\rangle}$ and $|e_2{\rangle}$,
and $\varepsilon_g$ is the energy of the common ground state $|g{\rangle}$.
In the following, we set $\varepsilon_{1}{\ge}\varepsilon_{2}$ and $\varepsilon_g=0$ for simplicity without losing any generality.
The Hamiltonian of three thermal baths is given by
$\hat{H}_b=\sum_{u=L,M,R}\hat{H}^u_b=\sum_{k,u}\omega_k\hat{a}^{\dag}_{k,u}\hat{a}_{k,u}$,
where $\hat{a}^{\dag}_{k,u}~(\hat{a}_{k,u})$ creates (annihilates) one phonon in the bath $u$ with frequency $\omega_k$.
The interaction between V-type system and the bath $L~(R)$ is described as
\begin{eqnarray}~\label{vu0}
\hat{V}_u=\sum_{k,i}
(g^i_{k,u}\hat{a}^{\dag}_{k,u}|g{\rangle}{\langle}e_i|+g^{i*}_{k,u}\hat{a}_{k,u}|e_i{\rangle}{\langle}g|),~u=L,R,
\end{eqnarray}
where $g^i_{k,u}$ is the coupling strength to emit one phonon into the bath $u$ by relaxing the V-type system from
$|e_i{\rangle}$ to $|g{\rangle}$, and $g^{i*}_{k,u}$ is the coupling strength in the reverse process.
It is easily to find that $\hat{V}_{L}$ and $\hat{V}_{R}$ can jointly participate in the transitions $|g{\rangle}\leftrightarrow|e_{1}{\rangle}$ and $|g{\rangle}\leftrightarrow|e_{2}{\rangle}$,
which may result in noised-induced coherence~\cite{swli2015aop}.
While the interaction $\hat{V}_M$ is given by
\begin{eqnarray}~\label{vm0}
\hat{V}_M=\sum_{k}
(g_{k,M}\hat{a}^{\dag}_{k,M}|e_2{\rangle}{\langle}e_1|+g^{*}_{k,M}\hat{a}_{k,M}|e_1{\rangle}{\langle}e_2|),
\end{eqnarray}
where $g_{k,M}~(g^*_{k,M})$ is the hopping strength from $|e_1{\rangle}$ to $|e_2{\rangle}$~(from $|e_2{\rangle}$ to $|e_1{\rangle}$) by
emitting (absorbing) one phonon into (from) the bath $M$.

\subsection{Redfield equation}
We consider the interaction between the V-type system and thermal baths (i.e., $\hat{V}_u~(u=L,M,R)$) is weak.
Based on the Born approximation, the whole density operator can be approximated as
$\hat{\rho}(t){\approx}\hat{\rho}_s(t){\otimes}(\Pi_u\hat{\rho}^u_b)$,
where $\hat{\rho}(t)$ is the density operator of the whole system, $\hat{\rho}_s(t)$ is the reduced density operator of the V-type system and $\hat{\rho}^u_b=\exp(-\hat{H}^u_b/(k_BT_u))/Z_u$
is the canonical distribution operator of the bath $u$, with the temperature of the $u$th bath $T_u$ and the partition function $Z_u=\textrm{Tr}_b\{\exp(-\hat{H}^u_b/(k_BT_u))\}$.
Moreover, we apply the Markovian approximation and  perturb $\hat{V}_u$ up to the second order, to obtain the quantum master equation as
\begin{eqnarray}~\label{re0}
\frac{d\hat{\rho}_s(t)}{dt}
&=&-i[\hat{H}_s,\hat{\rho}_s(t)]\\
&&+\frac{1}{2}\sum_{i,j;\sigma=\pm}\Gamma^{\sigma}_{ij}(\varepsilon_j)
([\hat{\phi}^\sigma_j\hat{\rho}_s(t),\hat{\phi}^{\overline{\sigma}}_i]+[\hat{\phi}^{{\sigma}}_i,\hat{\rho}_s(t)\hat{\phi}^{\overline\sigma}_j])\nonumber\\
&&+\frac{1}{2}\sum_{\sigma=\pm}\Gamma^{\sigma}_M(\Delta)
([\hat{\psi}^\sigma\hat{\rho}_s(t),\hat{\psi}^{\overline{\sigma}}]+[\hat{\psi}^{{\sigma}},\hat{\rho}_s(t)\hat{\psi}^{\overline{\sigma}}]),\nonumber
\end{eqnarray}
where $\hat{\phi}^+_i=|e_i{\rangle}{\langle}g|$ and $\hat{\phi}^-_i=|g{\rangle}{\langle}e_i|$ are transition operators between the ground state and the
$i$th excited state, $\hat{\psi}^+=|e_1{\rangle}{\langle}e_2|$ and $\hat{\psi}^-=|e_2{\rangle}{\langle}e_1|$ are transition operator between two excited states.
The energy bias between two excited states is $\Delta=\varepsilon_1-\varepsilon_2$.
The transition rates only involved with the left and right baths are $\Gamma^+_{ij}(\varepsilon_j)=\sum_{u=L,R}\gamma^u_{ij}(\varepsilon_j)n_u(\varepsilon_j)$
and $\Gamma^-_{ij}(\varepsilon_j)=\sum_{u=L,R}\gamma^u_{ij}(\varepsilon_j)(1+n_u(\varepsilon_j))$,
with the spectral function $\gamma^u_{ij}(\varepsilon_j)=\gamma^{u}_{ji}(\varepsilon_j)=2\pi\sum_kg^{i}_{k,u}g^{j*}_{k,u}\delta(\varepsilon_j-\omega_k)$
and the Bose-Einstein distribution function $n_u(\varepsilon_j)=1/[\exp(\varepsilon_j/(k_BT_u))-1]$.

$\Gamma^+_{ii}(\varepsilon_i)~(\Gamma^-_{ii}(\varepsilon_i))$ describes the particle transition rate
from the ground state probability $\rho_{gg}$ to the exited state $\rho_{ii}$ (from $\rho_{ii}$ to $\rho_{gg}$)
by absorbing (emitting) one phonon from (into) the left/right thermal bath.
While the rate $\Gamma^+_{12}(\varepsilon_i)~(\Gamma^-_{12}(\varepsilon_i))$ shows the noised-induced transition from the ground state probability $\rho_{gg}$ to the coherence term $\rho_{12}$(from $\rho_{12}$ to $\rho_{gg}$).
The noise induced coherence, also termed as Fano interference~\cite{kedorfman2013pnas}, has been extensively analyze in
steady state entanglement and energy transfer in biomolecular systems~\cite{dps2009pra,pnalbach2010njp},
quantum optics~\cite{vvkozlov2006pra} and quantum heat engines~\cite{kedorfman2013pnas,kedorfman2018pre}.
with the spectral function modulated in the regime $\gamma^u_{12}(\epsilon_j)\in[0,\sqrt{\gamma^u_{11}(\epsilon_j)\gamma^u_{22}(\epsilon_j)}]$~\cite{kedorfman2013pnas}.
Note that the coherence term $\rho_{12}$ is coupled with occupation probabilities $\rho_{ii}~(i=e_1,e_2,g)$
as $\Gamma^{\pm}_{12}(\varepsilon_i){\neq}0$.
Hence, the nonequilibrium quantum coherence may not only occur in the transient dynamics, but also persist in the steady state, which is termed as
nonequilibrium steady state coherence~\cite{swli2015aop,jjun2018scichina,zhwang2018njp}.

The transition rate involved with the middle bath is $\Gamma^{+}_M(\Delta)=\gamma_M(\Delta)n_M(\Delta)$
and $\Gamma^{-}_M(\Delta)=\gamma_M(\Delta)(1+n_M(\Delta))$,
with $\gamma_M(\Delta)=2\pi\sum_k|g_{k,M}|^2\delta(\Delta-\omega_k)$.
$\Gamma^{\pm}_{M}(\Delta)$ describes the probability transition between two exited states.
By including a third thermal bath, the molecular solar cell~\cite{moscully2011pnas} and quantum transistor~\cite{pbabd2014prl} have been extensively investigated within the three-terminal setup.
Particularly for quantum thermoelectric transistor within a double quantum dots device, the heat amplification was observed in the linear response regime~\cite{jhjiang2015prb}.
In the following, we will study the quantum thermal transistor in the V-type system far-from equilibrium.
In this paper, $\gamma^u_{ii}(\varepsilon_i)=\gamma^u_{ii}$ and $\gamma_{M}(\Delta)=\gamma_{M}$ are set constant for simplicity.
The extension of these spectral functions to frequency dependent is straightforward
(e.g., $\gamma^u_{ii}(\omega)=\alpha^{i}_u\omega\exp(-\omega/\omega_c)$), and will not qualitatively change the results.

\subsection{Nonequilibrium steady state coherence}

The nonequilibrium steady state has been revealed as a source to enhance the power and quantum efficiency
in the energy harvesting systems, where quantum coherence is unraveled to be crucial~\cite{moscully,moscully2011pnas,kedorfman2013pnas}.
Generally, the quantum coherence can be defined by the off-diagonal elements of the density matrix (i.e. $\rho_{12}$).
Here, following the same definition, we analyze the quantum coherence under the temperature bias at steady state.
From Eq.~(\ref{supre0}), it is easy to see that the diagonal elements $\rho_{ii}~(i=e_1,e_2,g)$ are dynamically coupled with the off-diagonal term
$\rho_{12}$. Hence, the quantum coherence may even appear after long time evolution.

\begin{figure}[tbp]
\includegraphics[scale=0.5]{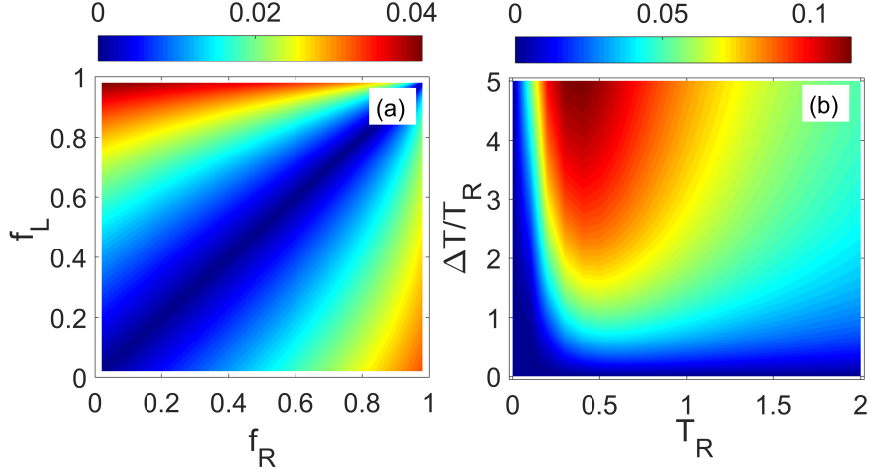}
\vspace{-1.0cm}
\caption{(Color online) Nonequilibrium steady state quantum coherence $|\rho^{ss}_{12}|$ within the two{\color{red}-}reservoir setup ($\gamma_M=0$) (a) by tuning noise-induced transition coefficients $\gamma^L_{12}$ and $\gamma^R_{12}$ with $T_L=2$ and $T_R=1$, and (b) by tuning the left and right temperatures $T_L$ and $T_R$ with ${\Delta}T=T_L-T_R$, $\gamma^L_{12}=\sqrt{\gamma^L_{11}\gamma^L_{22}}$ and $\gamma^R_{12}=0$.
The other system parameters are given by $\gamma^{L(R)}_{11}=\gamma^{L(R)}_{22}=0.01$ and $\varepsilon_L=\varepsilon_R=1$.}~\label{fig2}
\end{figure}


The analytical expression of the steady state quantum coherence is obtained at certain specific case.
It is generally quite difficult to obtain the analytical expression of the quantum coherence.
However, at resonance ($\varepsilon_1=\varepsilon_2=\varepsilon$) and without the middle reservoir ($\gamma_M=0$),
the steady state coherence is given by (see Eq.~\ref{suprho12})
\begin{eqnarray}
\rho^{ss}_{12}=\frac{\Gamma^-_{11}(\varepsilon)\Gamma^-_{22}(\varepsilon)\Gamma^-_{12}(\varepsilon)}{[\Gamma^-_{11}(\varepsilon)+\Gamma^-_{22}(\varepsilon)]A}
\left[\frac{2\Gamma^+_{12}(\varepsilon)}{\Gamma^-_{12}(\varepsilon)}-\frac{\Gamma^+_{11}(\varepsilon)}{\Gamma^-_{11}(\varepsilon)}-\frac{\Gamma^+_{22}(\varepsilon)}{\Gamma^-_{22}(\varepsilon)}\right],
\end{eqnarray}
where $A=\Gamma^-_{11}(\varepsilon)[\Gamma^-_{22}(\varepsilon)+\Gamma^+_{22}(\varepsilon)]+\Gamma^+_{11}(\varepsilon)\Gamma^-_{22}(\varepsilon)
-\Gamma^-_{12}(\varepsilon)[\Gamma^-_{12}(\varepsilon)+2\Gamma^+_{12}(\varepsilon)]$.
It needs to point out that the steady state coherence here is completely induced by the noise-induced interference,
and is irrelevant with the concept of decoherence free subspace~\cite{lmduan1997prl,pzanardi1997prl,dalidar1998prl}.

As known from Eq.~(\ref{re0}), the noise-induced interference, quantified by $\Gamma^+_{12}(\varepsilon_i)$, is irrelevant with the direct hopping assisted by the middle reservoir.
Hence, we include the two-reservoir setup to study the quantum coherence by setting $\gamma_M=0$, shown at Fig.~\ref{fig2}(a).
As $\gamma^L_{12}=\gamma^R_{12}$, the steady state coherence shows globally minimal ($\rho^{ss}_{12}=0$). This is consistent with the vanishing condition of the quantum coherence at Eq.~(\ref{cond2}).
While as  $\gamma^L_{12}{\neq}\gamma^R_{12}$, quantum coherence shows monotonic enhancement by increasing the bias of noise-induced transition coefficient $|\gamma^L_{12}-\gamma^R_{12}|$, and exhibits maximum at $\gamma^L_{12}=\sqrt{\gamma^L_{11}\gamma^L_{22}}$ and $\gamma^R_{12}=0$.
Though not shown here, quantum coherence still sustains under the off-resonance case $\varepsilon_1{\neq}\varepsilon_2$.
The temperature dependence of the quantum coherence with large noise-induced transition coefficient bias ($\gamma^L_{12}=\sqrt{\gamma^L_{11}\gamma^L_{22}}$ and $\gamma^R_{12}=0$) is plotted in Fig.~\ref{fig2}(b).
It is found that in the moderate temperature  regime (e.g., $T_R=0.5$), quantum coherence is dramatically enhanced by increasing the temperature bias ($T_L-T_R$).
While in the low and high temperature regimes, $\rho^{ss}_{12}$ becomes small but still nonzero.
Hence, there exists an optimal temperature regime to generate the comparatively large steady state quantum coherence.

\section{Heat current fluctuations}
We introduce a full counting statistics (FCS) method to count the energy flow into thermal baths~\cite{mesposito2009rmp,llevitov1993jetp,mcampisi2011rmp},
which is powerful to detect the crucial information of heat current fluctuations, and encoded in the corresponding cumulant generating function.
Specifically, we introduce a counting field set $\{\chi\}=\{\chi_L,\chi_R\}$ to the Hamiltonian as
$\hat{H}_{\{\chi\}}=e^{i\sum_{u=L,R}\chi_u\hat{H}_v/2}\hat{H}e^{-i\sum_{u=L,R}\chi_u\hat{H}_u/2}=
\hat{H}_s+\hat{H}_b+\hat{V}_M+\sum_{u=L,R}\hat{V}_u({\chi_u})$~\cite{mesposito2009rmp},
with $\chi_u$ the counting field parameter of the $u$th bath.
The modified system-bath interaction is expressed as
\begin{eqnarray}~\label{vu1}
\hat{V}_u(\chi_u)=\sum_{k,i}(g^i_{k,v}e^{i\omega_k\chi_u}\hat{a}^{\dag}_{k,u}|g{\rangle}{\langle}e_i|
+g^{i*}_{k,u}e^{-i\omega_k\chi_u}\hat{a}_{k,u}|e_i{\rangle}{\langle}g|).
\end{eqnarray}
Based on the Born-Markov approximation, we perturb the interaction Eq.~(\ref{vu1}) up to the second order,
and obtain the modified quantum master equation  (see details at appendix B)
\begin{eqnarray}~\label{re1}
\frac{d\hat{\rho}_{\{\chi\}}}{dt}&=&-i[\hat{H}_s,\hat{\rho}_{\{\chi\}}]\\
&&-\frac{1}{2}\sum_{i,j;\sigma=\pm}\Gamma^{\sigma}_{ij}(\varepsilon_j)(\hat{\phi}^{\overline{\sigma}}_i\hat{\phi}^{{\sigma}}_j\hat{\rho}_{\{\chi\}}
+\hat{\rho}_{\{\chi\}}\hat{\phi}^{\overline{\sigma}}_j\hat{\phi}^{{\sigma}}_i)\nonumber\\
&&+\frac{1}{2}\sum_{i,j;\sigma=\pm}(\Gamma^{\sigma}_{ij}(\varepsilon_i,\{\chi\})+\Gamma^{\sigma}_{ij}(\varepsilon_j,\{\chi\}))
\hat{\phi}^\sigma_i\hat{\rho}_{\{\chi\}}\hat{\phi}^{\overline{\sigma}}_j\nonumber\\
&&+\frac{1}{2}\sum_{\sigma=\pm}\Gamma^{\sigma}_m(\Delta)
([\hat{\psi}^\sigma\hat{\rho}_{\{\chi\}},\hat{\psi}^{\overline{\sigma}}]+[\hat{\psi}^{{\sigma}},\hat{\rho}_{\{\chi\}}\hat{\psi}^{\overline{\sigma}}]),\nonumber
\end{eqnarray}
where the modified transition rates are
$\Gamma^+_{ij}(\omega,\{\chi\})=\sum_v\gamma^v_{ij}n_v(\omega)e^{-i\omega\chi_v}$ and
$\Gamma^-_{ij}(\omega,\{\chi\})=\sum_v\gamma^v_{ij}(1+n_v(\omega))e^{i\omega\chi_v}$.
In absence of the counting fields ($\chi_L=\chi_R=0$),
this modified quantum master equation returns back to the standard version at Eq.~(\ref{re0}).


From the definition at Eq.~(\ref{supju1}), the steady state heat current into the right bath is given by
\begin{eqnarray}~\label{jer}
J^e_R&=&\sum_{j=1,2}\varepsilon_j\gamma^R_{jj}[(1+n_R(\varepsilon_j))\rho^{ss}_{jj}-n_R(\varepsilon_j)\rho^{ss}_{gg}]\\
&&+\frac{1}{2}\sum_{j=1,2}\varepsilon_j\gamma^R_{12}(1+n_R(\varepsilon_j))(\rho^{ss}_{12}+\rho^{ss}_{21}).\nonumber
\end{eqnarray}
The first term on the right side shows the population transfer process between the excited state population ($\rho^{ss}_{jj}$) and the ground state population ($\rho^{ss}_{gg}$).
The second term denotes the contribution of the noise-induced interference to the steady state heat transfer, which is quantified by $\gamma^L_{12}$.
Similarly, the heat current into the left bath is
\begin{eqnarray}~\label{jel}
J^e_L&=&\sum_{j=1,2}\varepsilon_j\gamma^L_{jj}[(1+n_L(\varepsilon_j))\rho^{ss}_{jj}-n_L(\varepsilon_j)\rho^{ss}_{gg}]\\
&&+\frac{1}{2}\sum_{j=1,2}\varepsilon_j\gamma^L_{12}(1+n_L(\varepsilon_j))(\rho^{ss}_{12}+\rho^{ss}_{21}),\nonumber
\end{eqnarray}
and the heat current into the middle thermal bath is
\begin{eqnarray}~\label{jem}
J^e_M=\Delta\gamma_M[(1+n_M(\Delta))\rho^{ss}_{11}-n_M(\Delta)\rho^{ss}_{22}].
\end{eqnarray}
For $J^e_M$ involve the heat exchange between two excited states, which excludes the noise-induced interference induced transfer process.
They fulfill the energy conservation law as $J^e_L+J^e_R+J^e_M=0$.

\begin{figure}[tbp]
\includegraphics[scale=0.5]{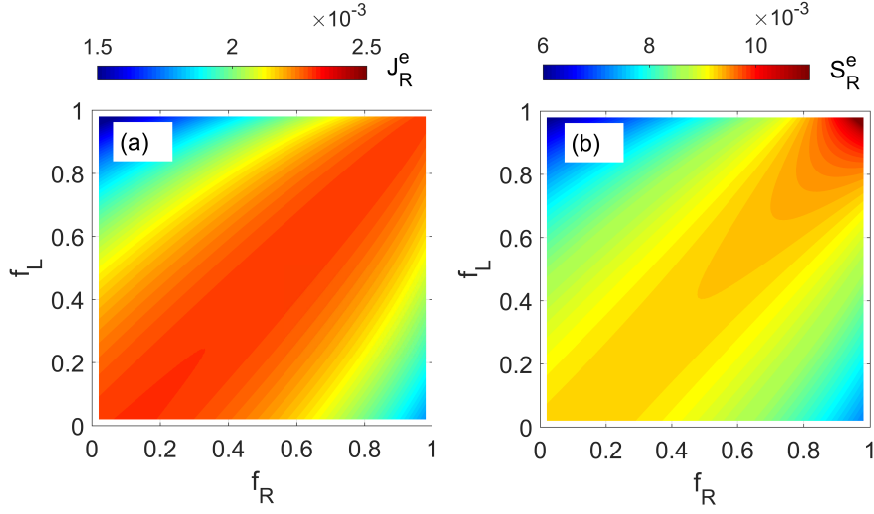}
\vspace{-0.5cm}
\caption{(Color online) (a) Steady state heat flux and (b)  noise power into the right thermal bath, by modulating the cross-induced transition coefficients
$\gamma^L_{12}$ and $\gamma^R_{12}$.  The other system parameters are given by $T_L=2$, $T_R=1$, $\gamma^{L(R)}_{11}=\gamma^{L(R)}_{22}=0.01$, $\gamma_M=0$ and $\varepsilon_L=\varepsilon_R=1$.}~\label{fig21}
\end{figure}

Then, we investigate the effect of the noise-induced interference on heat currents cumulants (e.g., $J^e_R$ and $S^e_{RR}$, see Eq.(B5)) with $\gamma_M=0$, shown at Fig.~\ref{fig21}.
For the heat current into the right bath at Fig.~\ref{fig21}(a), it is found that $J^e_R$ is  dramatically suppressed by the increase of  the bias of noise-induced transition coefficients ($|\gamma^L_{12}-\gamma^R_{12}|$),
and becomes minimum at the limiting regimes (e.g., $\gamma^L_{12}=\sqrt{\gamma^L_{11}\gamma^L_{22}}$, $\gamma^R_{12}=0$).
While the noise power characterizes the correlations of currents, which originate from stochastic processes in the nonequilibrium transfer~\cite{mesposito2009rmp}.
It is shown that for the condition $\gamma^L_{12}=\gamma^R_{12}$, $S^e_{RR}$ shows monotonic enhancement by increasing $\gamma^L_{12}$, and becomes maximum at $\gamma^L_{12}=\gamma^R_{12}=1$, which is exhibited in Fig.~\ref{fig21}(b).
Whereas $S^e_{RR}$ is strongly suppressed at large bias of $|\gamma^L_{12}-\gamma^R_{12}|$.
Hence, the large bias of noise-induced transition coefficients $|\gamma^L_{12}-\gamma^R_{12}|$ deteriorates the heat current and the noise power.

\begin{figure}[tbp]
\includegraphics[scale=0.5]{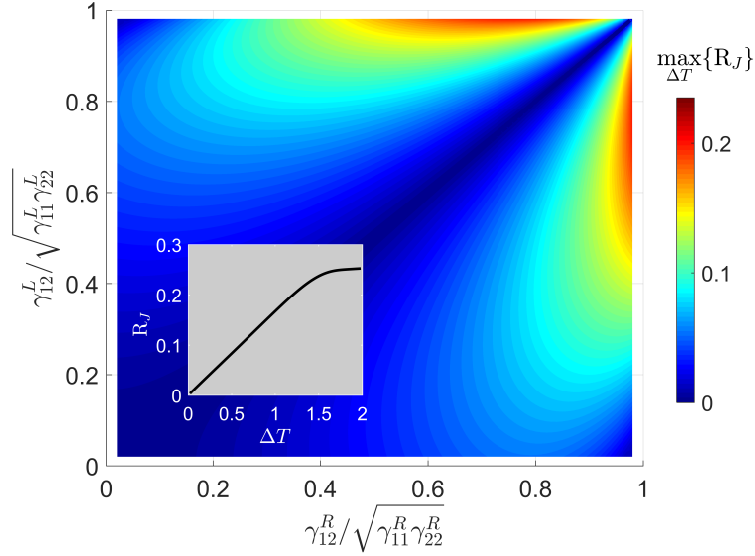}
\caption{(Color online) Maximal rectification factor of the heat flux $\max_{{\Delta}T}\{\textrm{R}_J\}$ by tuning the temperature bias ${\Delta}T$ as a function of the noise-induced coefficient $\gamma^{L(R)}_{12}$.
The inset shows the behavior of $\textrm{R}_J$ as a function of ${\Delta}T$ with the coefficients $\gamma^{L}_{12}=0.8\sqrt{\gamma^{L}_{11}\gamma^{L}_{22}}$
and $\gamma^{R}_{12}=\sqrt{\gamma^{R}_{11}\gamma^{R}_{22}}$.
The bath temperatures are $T_L=T_0+{\Delta}T/2$, $T_R=T_0+{\Delta}T/2$ and $T_0=1$.
The other system parameters are given by $\gamma^{L(R)}_{11}=\gamma^{L(R)}_{22}=0.01$ and $\varepsilon_L=\varepsilon_R=1$.}~\label{fig3}
\end{figure}

\section{Noise-induced thermal rectification}
Thermal rectification effect has been extensively investigated in electronics and phononics, which typically is a two-terminal phenomena~\cite{bwli2004prl,dsegal2005prl,nbli2010rmp,fgiazotto2015nn}.
Several typical definitions of the thermal rectification have been proposed, e.g.,
the rectifier ratio $|J^e_R(T_0,\Delta{T})/J^e_R(T_0,-\Delta{T})|$~\cite{bwli2004prl,lawu2009prl} with $T_L=T_0-{\Delta}T/2$
and $T_R=T_0+{\Delta}T/2$,
and the rectification efficiency~\cite{dsegal2005prl,lfzhang2010prb}
\begin{eqnarray}~\label{rj1}
\textrm{R}_J=\frac{|J^e_R(T_0,\Delta{T})+J^e_R(T_0,-\Delta{T})|}{\max\{J^e_R(T_0,\Delta{T}),-J^e_R(T_0,-\Delta{T})\}}
\end{eqnarray}
These definitions capture the asymmetric behavior of the heat flux by interchanging the temperatures of two baths.
In this paper, we select Eq.~(\ref{rj1}) to quantify the thermal rectification with $J^e_R$ at Eq.~(\ref{jer}),
i.e., the rectification occurs as $\textrm{R}_J>0$.

We analyze how the noise-induced interference generates steady state thermal rectification of the current within the two-terminal setup ($\gamma_M=0$).
Under the condition $\varepsilon_j=\varepsilon$ and $\gamma^{L(R)}_{jj}=\gamma~(j=1,2)$,
the heat current is expressed as (see Eq.~(\ref{jer}) and Eq.~(\ref{suprho12}))
\begin{eqnarray}
J^e_R&=&\frac{2\varepsilon\gamma}{\mathcal{A}}(\Gamma^-\gamma-\Gamma^-_{12}\gamma^L_{12})[n_L(\varepsilon)-n_R(\varepsilon)]\\
&&+\frac{2\varepsilon\gamma\gamma^R_{12}(\gamma^L_{12}-\gamma^R_{12})}{\mathcal{A}}[1+n_R(\varepsilon)][n_L(\varepsilon)-n_R(\varepsilon)],\nonumber
\end{eqnarray}
with the coefficient
$\mathcal{A}=(\Gamma^-)^2+2\Gamma^+\Gamma^--\Gamma^-_{12}(2\Gamma^+_{12}+\Gamma^-_{12})$
and
the rates $\Gamma^+=\gamma[n_L(\varepsilon)+n_R(\varepsilon)]$, $\Gamma^-=\gamma[2+n_L(\varepsilon)+n_R(\varepsilon)]$,
$\Gamma^+_{12}=[\gamma^L_{12}n_L(\varepsilon)+\gamma^R_{12}n_R(\varepsilon)]$ and
$\Gamma^-_{12}=\gamma^L_{12}[1+n_L(\varepsilon)]+\gamma^R_{12}[1+n_R(\varepsilon)]$.
For $J^e_R$, the first term on the right side comes from the contributions of populations ($\rho^{ss}_{jj}~(j=1,2,g)$),
and the second term is contributed by the steady state coherence ($\rho^{ss}_{12}$).
When $\gamma^{L}_{12}{\neq}\gamma^{R}_{12}$, it is interesting to find that both two components of $J^e_R$ show the nonreciprocal relationship,
for $\gamma^L_{12}n_L(\varepsilon)+\gamma^R_{12}n_R(\varepsilon){\neq}\gamma^L_{12}n_R(\varepsilon)+\gamma^L_{12}n_R(\varepsilon)$ by exchanging $T_L$ with $T_R$.
Thus, it clearly shows the thermal rectification feature.
While as $\gamma^{L}_{12}{=}\gamma^{R}_{12}$, the contribution from the steady state coherence vanishes.
Moreover, the rates $\Gamma^{\pm}_{12}$ are simplified as
$\Gamma^+_{12}=\gamma_{12}[n_L(\varepsilon)+n_R(\varepsilon)]$ and
$\Gamma^-_{12}=\gamma_{12}[2+n_L(\varepsilon)+n_R(\varepsilon)]$,
which both become invariant by exchanging two bath temperatures $T_L$ and $T_R$.
The current is simplified as $J_R=2\gamma\varepsilon[n_L(\varepsilon)-n_R(\varepsilon)]/[2+3n_L(\varepsilon)+3n_R(\varepsilon)]$,
and the thermal rectification behavior naturally disappears.
Therefore, we conclude that the bias of noise-induced interference(i.e. $\gamma^{L}_{12}{\neq}\gamma^{R}_{12}$)
is the origin to exhibit the thermal rectification, i.e. $\textrm{R}_{J}{\neq}0$.

Furthermore, we study the maximal rectification factor $\max_{{\Delta}T}\{\textrm{R}_J\}$  under the influence of noise-induced coefficient $\gamma^{L(R)}_{12}$ in Fig.~\ref{fig3}, where $\max_{{\Delta}T}\{\textrm{R}_J\}$ is the maximal value of $\textrm{R}_J$ by tuning the temperature bias ${\Delta}T$ for given $\gamma^{L(R)}_{12}$.
Interestingly, the rectification factor is significantly enhanced in the optimal coefficient regime(e.g., $\gamma^{L}_{12}/\sqrt{\gamma^{L}_{11}\gamma^{L}_{22}}{\approx}0.76$ and $\gamma^{R}_{12}/\sqrt{\gamma^{R}_{11}\gamma^{R}_{22}}{\approx}0.98$),
rather than shows monotonic increase with the bias $|\gamma^{L}_{12}-\gamma^{R}_{12}|$.
Then, we investigate the behavior of $\textrm{R}_J$ by tuning the temperature bias in the optimal coefficient regime in the inset of Fig.~\ref{fig3}.
It is found that ${\Delta}T$ monotonically enhances the rectification factor.
Hence, the heat rectification factor favors the large temperature bias.

In previous works of the  thermal rectification, the sufficient condition for the appearance of  heat rectification has been analyzed in two-reservoir spin-boson model and boson-boson model~\cite{lawu2009prl}.
Consequently, such condition has also been analyzed in the Z-type three level-system~\cite{lawu2009pre}, in which there is no noise-induced interference.
Asymmetric structures of the quantum system and system-bath interaction both is found to contribute to the quantum rectification~\cite{lawu2009prl}.
However, the influence of the noise-induced interference on the rectification effect is lack of exploitation in three-level system.
In this work, we clearly indicate that in V-type system the noise-induced interference is able to exhibit the thermal rectification.
It should be noted that for the nonequilibrium $\Lambda$-type system, though not shown here,
the feature of noise-induced interference induced thermal rectification can also be observed,
which is mainly due to the similar structure of the system-bath interaction compared to the V-type system~\cite{swli2015aop}.

Moreover, the previous sufficient condition in Ref.~\cite{lawu2009prl} can be recovered based on the expression of heat flux at Eq.~(\ref{jer}),
in absence of the noise-induced  interference (i.e. $\gamma^L_{12}=\gamma^R_{12}=0$).
Specifically, we obtain the corresponding expression of heat current into the right bath as
\begin{eqnarray}
J_R&=&\frac{\gamma^L_{11}\gamma^R_{11}}{A^{\prime}}\Gamma^-_{22}(\varepsilon_2)[n_L(\varepsilon_1)-n_R(\varepsilon_1)]\varepsilon_1\nonumber\\
&&+\frac{\gamma^L_{22}\gamma^R_{22}}{A^{\prime}}\Gamma^-_{11}(\varepsilon_2)[n_L(\varepsilon_2)-n_R(\varepsilon_2)]\varepsilon_2\nonumber
\end{eqnarray}
with the coefficient $A^{\prime}=\Gamma^-_{11}(\varepsilon_1)(\Gamma^+_{22}(\varepsilon_2)+\Gamma^-_{22}(\varepsilon_2))+\Gamma^+_{11}(\varepsilon_1)\Gamma^-_{22}(\varepsilon_2)$.
Then by setting $\varepsilon_2=0$, it results in $\Gamma^{+}_{22}(\varepsilon_2){=}\Gamma^{-}_{22}(\varepsilon_2)$.
The current is reduced to
$J_R={\gamma^L_{11}\gamma^R_{11}[n_L(\varepsilon_1)-n_R(\varepsilon_1)]\varepsilon_1}/[{\gamma^L_{11}(2+3n_L(\varepsilon_1))+\gamma^R_{11}(2+3n_R(\varepsilon_1))}]$.
Finally, the condition is recovered as
\begin{eqnarray}
{n_L(\varepsilon_1)}/{\gamma^L_{11}}-{n_L(\varepsilon_1)}/{\gamma^R_{11}}
={n_R(\varepsilon_1)}/{\gamma^L_{11}}-{n_R(\varepsilon_1)}/{\gamma^R_{11}},
\end{eqnarray}
which becomes identical with the key result in Ref.~\cite{lawu2009prl}.

\begin{figure}[tbp]
\includegraphics[scale=0.4]{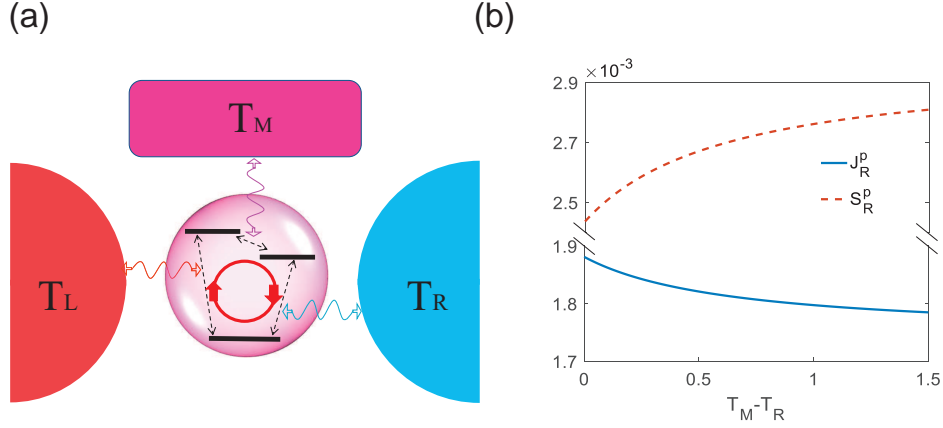}
\caption{(Color online) (a) Schematic diagram of heat flows (red circle) cooperatively contributed by three reservoirs,
and (b) heat current and noise power by tuning the temperature of middle bath $T_M$ with $\gamma^L_{22}=\gamma^R_{11}=0$ and $\gamma^L_{12}=\gamma^R_{12}=0$.
The other system parameters are $\varepsilon_1=1.1$, $\varepsilon_2=0.9$, $\gamma^{L}_{11}=\gamma^{R}_{22}=\gamma_M=0.01$,
$T_L=2$, and $T_R=0.5$.}~\label{fig4}
\end{figure}

\begin{figure}[tbp]
\includegraphics[scale=0.55]{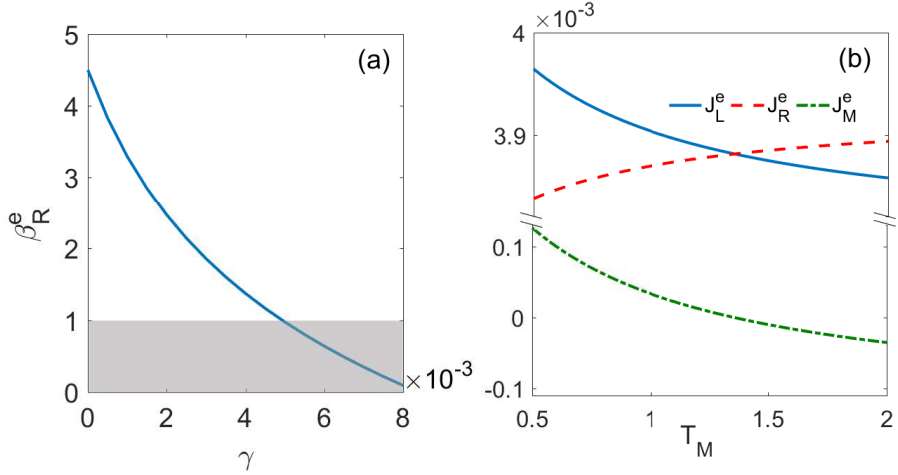}
\vspace{-1.5cm}
\caption{(Color online) (a) Heat amplification factor by tuning the coupling strength $\gamma^L_{22}=\gamma^R_{11}=\gamma$,
and (b) heat currents by tuning the temperature of middle thermal bath with $\gamma^L_{22}=\gamma^R_{11}=0.01$.
The other system parameters are given by
$\varepsilon_1=1.1$, $\varepsilon_2=0.9$, $\gamma^L_{11}=\gamma^R_{22}=\gamma_M=0.01$, $\gamma^L_{12}=\gamma^R_{12}=0$,
$T_L=2$, and $T_R=0.5$. }~\label{fig5}
\end{figure}

\section{Heat amplification}

Heat amplification is the key component to realize quantum thermal transistors within three-terminal setups~\cite{bli2006apl,nbli2010rmp}.
The amplification factor is defined by the ratio of the change of the current $J^{e}_L(J^{e}_R)$ on the change of the middle bath current
$J^{e}_M$
\begin{eqnarray}~\label{betaeu1}
\beta^{e}_u=|{\partial}J^{e}_u/{\partial}J^{e}_M|,~u=L,R.
\end{eqnarray}
According to the energy conservation relationship $\sum_{u=L,M,R}J^e_u=0$,
the amplification factor $\beta^e_R$ can be re-expressed as
\begin{eqnarray}~\label{relation1}
\beta^e_R=|\beta^e_L+(-1)^\theta|,
\end{eqnarray}
with $\theta=0$ for ${\partial}J^{e}_L/{\partial}J^{e}_M>0$
and
$\theta=1$ for ${\partial}J^{e}_L/{\partial}J^{e}_M<0$.
Traditionally, the amplification effect occurs once $\beta^e_{L(R)}>1$.

To simplify the analysis of heat amplification in the nonequilibrium V-type system, we first ignore the transition between states $|e_{2(1)}{\rangle}$
and $|g{\rangle}$ mediated by the left(right) bath($\gamma^L_{22}=\gamma^R_{11}=0$),
and the noise-induced interference($\gamma^{L(R)}_{12}=0$).
Thus, the particle current Eq.~(\ref{jp1}) into the right bath based on FCS is expressed as
\begin{eqnarray}
J^p_R=(\Gamma^{+}_{11}\Gamma^-_M\Gamma^-_{22}-\Gamma^+_{22}\Gamma^+_M\Gamma^{-}_{11})/\mathcal{B},
\end{eqnarray}
with the coefficient $\mathcal{B}=(\Gamma^+_{22}+\Gamma^-_{22}+\Gamma^+_M)(\Gamma^+_{11}+\Gamma^-_{11}+\Gamma^-_M)
-(\Gamma^-_M-\Gamma^+_{22})(\Gamma^+_M-\Gamma^+_{11})$,
$\Gamma^{+}_{11(22)}=\gamma^{L(R)}_{11(22)}n_{L(R)}(\varepsilon_{1(2)})$,
$\Gamma^{-}_{11(22)}=\gamma^{L(R)}_{11(22)}[1+n_{L(R)}(\varepsilon_{1(2)})]$,
$\Gamma^+_M=\gamma_{M}n_{M}(\varepsilon_{1}-\varepsilon_2)$
and
$\Gamma^-_M=\gamma_{M}[1+n_{M}(\varepsilon_{1}-\varepsilon_2)]$.
From $J^p_R$, it it known that the carrier needs finish a cyclic flow to make the steady state heat current in Fig.~\ref{fig4}(a).
In a microscopic view, one carrier should be excited from the ground state to $|e_1{\rangle}$ by absorbing one phnon from the left bath, then
transferred to the excited state $|e_2{\rangle}$ via the middle bath and finally relaxed to the ground state again
via emitting one phonon into the right bath.
Based on this picture, it is interesting to find the particle currents into the left and middle baths are given by
$J^p_L=-J^p_R$ and $J^p_M=J^p_R$, which implies the magnitudes of particle currents are the same.
Consequently, heat fluxes into three reservoirs are straightforwardly obtained by
$J^e_R=\varepsilon_2J^p_R$, $J^e_L=-\varepsilon_1J^p_R$,
and $J^e_M=(\varepsilon_1{-}\varepsilon_2)J^p_R$, respectively.
Then, the heat amplification factor from Eq.~(\ref{betaeu1}) is specified as
\begin{eqnarray}
\beta^e_{R}=\left|\frac{\varepsilon_2({\partial}J^p_R/{\partial}T_M)}{{\Delta}({\partial}J^p_M/{\partial}T_M)}\right|
=\left|\frac{\varepsilon_2}{\varepsilon_1-\varepsilon_2}\right|.
\end{eqnarray}
The heat amplification effect can be observed once $|\varepsilon_2/(\varepsilon_1-\varepsilon_2)|>1$,
and becomes apparent as $|\varepsilon_1{-}\varepsilon_2|{\ll}{\varepsilon_2}$.
It should be pointed out that ${\varepsilon_2}$ will never equal ${\varepsilon_1}$,
for the energy exchange through the excited state transitions
should be accompanied by the finite energy exchange of the system with the middle bath($\varepsilon_1{\neq}\varepsilon_2$).

In previous works of quantum thermal transistor, negative differential thermal conductance was believed to be a compulsory ingredient to realize the heat amplification~\cite{bli2006apl,nbli2010rmp,kjoulain2016prl,cwang2018pra,gtcraven2017prl,pbabd2014prl,kjoulain2015apl}.
Within the three-terminal setup, the NDTC generally occurs as the temperature bias $|T_v-T_M|~(v=L,R)$ increases,
the heat current is suppressed~\cite{nbli2010rmp,cwang2018pra}.
Recently, within the linear response regime, J. H. Jiang {{\itshape}{et al}.} found the heat amplification in gate-tunable double quantum dots without
negative differential thermal conductance~\cite{jhjiang2015prb}.
Here, our result clearly shows that the heat amplification effect can also be realized far-from equilibrium (finite temperature bias),
in absence of the negative differential thermal conductance.
Moreover, the heat current shows monotonic decrease with the increase of the middle bath temperature $T_M$,
whereas the  noise power $S^e_R$ exhibits enhancement (see Fig.~\ref{fig4}(b)).
Hence, it is proper to observe the heat amplification effect in comparatively low temperature regime (e.g., $T_M{\approx}0.5$), with high signal to noise ratio.

Next, we tune on $\gamma^L_{22}=\gamma^R_{11}=\gamma$ to analyze the influence of the two-terminal process on the heat amplification in Fig.~\ref{fig5}(a),
which includes direction transition between the left and right baths.
To quantify the heat amplification, we apply the maximum of the amplification factor by modulating $T_{M}$ as
\begin{eqnarray}
\beta^e_{R,\max}=\max_{\{T_M\}}\{\beta^e_R\}
=\left|\frac{\varepsilon_2}{\varepsilon_1-\varepsilon_2}\right|{\times}\max_{\{T_M\}}
\left\{\left|\frac{({\partial}J^p_R/{\partial}T_M)}{({\partial}J^p_M/{\partial}T_M)}\right|\right\}.
\end{eqnarray}
It is found that the heat amplification factor $\beta^e_{R,\max}$ decreases gradually by increasing $\gamma$, and finally drops below one (e.g., $\gamma=0.006$).
To see this clearly, we study the behavior of heat currents at $\gamma=0.01$ by tuning $T_M$ at Fig.~\ref{fig5}(b).
The change of $J^e_R$ is much smaller than the change of $J^e_M$, which results in $\beta^e_{R,\max}{\ll}1$.
Similarly, from the relation of the amplification factors at Eq.~(\ref{relation1}),
it is known that $\beta^e_{L,\max}{\approx}1$.
Hence, we conclude that two-terminal transport process is detrimental to the generation of the heat amplification.

\section{Conclusion}
To give a brief summary,
we investigate the quantum heat transfer in a nonequilibrium V-type system with weak system-bath interactions by applying the Redfield master equation.
The nonequilibrium quantum coherence is analytically investigated at steady state,
and the coherence can be optimized by tuning bath temperatures.
Within the two-bath setup,
the finite bias of noise-induced transition coefficients $|\gamma^L_{12}-\gamma^R_{12}|$ is found to enhance the steady state quantum coherence.
While for the noise power, it becomes maximal with largest noise-induced interference for both thermal baths.
Moreover, it is interesting to find that the noise-induced interference may rectify heat currents, which provides a new scheme of thermal rectification.
This clearly answer the first question in the introduction section.
Within  the three-bath setup,  even in absence of the negative differential thermal conductance,
a giant amplification factor is analytically obtained far-from equilibrium, which tightly relies on the energy levels of the excited states,
which is the answer for the second question raised in the introduction section.
Hence, this provides a smart way to control the amplification effect by modulating the V-type system structure.

\section{Acknowledgements}
C.W. is supported by the National Natural Science Foundation of China under Grant No. 11704093.
D.Z.X. is supported by the National Natural Science Foundation of China under Grant No. 11705008 and Beijing Institute of Technology Research
Fund Program for Young Scholars.
X.L.G. acknowledges support by the National Natural Science Foundation of China under Grant No. 11374266.

\appendix

\section{Steady state populations}
Following Eq.~(\ref{re0}), the dynamical equation of V-type system is given by
\begin{eqnarray}~\label{supre0}
\frac{\partial\rho_{11}}{\partial{t}}&=&-(\Gamma^-_{11}(\varepsilon_1)+\Gamma^-_M(\Delta))\rho_{11}+\Gamma^+_M(\Delta)\rho_{22}+\Gamma^+_{11}(\varepsilon_1)\rho_{gg}
-\frac{1}{2}\Gamma^-_{12}(\varepsilon_2)(\rho_{12}+\rho_{21}),\\
\frac{\partial\rho_{22}}{\partial{t}}&=&\Gamma^-_M(\Delta)\rho_{11}-(\Gamma^-_{22}(\varepsilon_2)+\Gamma^+_M(\Delta))\rho_{22}+\Gamma^+_{22}(\varepsilon_2)\rho_{gg}
-\frac{1}{2}\Gamma^-_{12}(\varepsilon_1)(\rho_{12}+\rho_{21}),\nonumber\\
\frac{\partial\rho_{gg}}{\partial{t}}&=&-(\Gamma^+_{11}(\varepsilon_1)+\Gamma^+_{22}(\varepsilon_2))\rho_{gg}+\Gamma^-_{11}(\varepsilon_1)\rho_{11}+
\Gamma^-_{22}(\varepsilon_2)\rho_{22}+\frac{1}{2}(\Gamma^-_{12}(\varepsilon_1)+\Gamma^-_{12}(\varepsilon_2))(\rho_{12}+\rho_{21}),\nonumber\\
\frac{\partial\rho_{12}}{\partial{t}}&=&-i\Delta\rho_{12}-\frac{1}{2}(\Gamma^-_{11}(\varepsilon_1)+\Gamma^-_{22}(\varepsilon_2))\rho_{12}
-\frac{1}{2}(\Gamma^-_{12}(\varepsilon_1)\rho_{11}+\Gamma^-_{12}(\varepsilon_2)\rho_{22})\nonumber\\
&&+\frac{1}{2}(\Gamma^+_{12}(\varepsilon_1)+\Gamma^+_{12}(\varepsilon_2))\rho_{gg}
-\frac{1}{2}(\Gamma^+_{M}(\Delta)+\Gamma^-_{M}(\Delta))\rho_{12},\nonumber\\
\frac{\partial\rho_{21}}{\partial{t}}&=&-i\Delta\rho_{21}-\frac{1}{2}(\Gamma^-_{11}(\varepsilon_1)+\Gamma^-_{22}(\varepsilon_2))\rho_{21}
-\frac{1}{2}(\Gamma^-_{12}(\varepsilon_1)\rho_{11}+\Gamma^-_{12}(\varepsilon_2)\rho_{22})\nonumber\\
&&+\frac{1}{2}(\Gamma^+_{12}(\varepsilon_1)+\Gamma^+_{12}(\varepsilon_2))\rho_{gg}
-\frac{1}{2}(\Gamma^+_{M}(\Delta)+\Gamma^-_{M}(\Delta))\rho_{21}.\nonumber
\end{eqnarray}

We would like to know at what condition the quantum coherence will become zero at steady state.
From the dynamical equation Eq.~(\ref{supre0}), the condition for the vanishing of steady state quantum coherence ($\rho^{ss}_{12}=0$) is given by
\begin{eqnarray}~\label{cond1}
&&\sum_{j=1,2}\Gamma^-_{12}(\varepsilon_j)\left[\frac{\Gamma^+_{jj}(\varepsilon_j)}{\Gamma^-_{jj}(\varepsilon_j)}
-(1+\frac{\Gamma^+_{M}(\Delta)}{\Gamma^-_{22}(\varepsilon_2)}+\frac{\Gamma^-_{M}(\Delta)}{\Gamma^-_{11}(\varepsilon_1)})\frac{\Gamma^+_{12}(\varepsilon_j)}{\Gamma^-_{12}(\varepsilon_j)}\right]\nonumber\\
&&[\Gamma^+_{11}(\varepsilon_1)+\Gamma^+_{22}(\varepsilon_2)]
\frac{\Gamma^-_{12}(\varepsilon_{1})\Gamma^+_M(\Delta)+\Gamma^-_{12}(\varepsilon_{2})\Gamma^-_M(\Delta)}{\Gamma^-_{11}(\varepsilon_1)\Gamma^-_{2}(\varepsilon_2)}=0,
\end{eqnarray}
which is cooperatively contributed by three baths.
Therefore, we conclude that the steady state could be zero even with the noise-induced interference ($\gamma^{L(R)}_{12}{\neq}0$).
This fact provides some insight to investigate the effect of noise-induced coherence on the efficiency bound in quantum heat engine.
As $\Gamma^{\pm}_M(\Delta)=0$~($\gamma_M=0$), the condition is reduced to the two-reservoir case
\begin{eqnarray}~\label{cond2}
\Gamma^-_{12}(\varepsilon_1)\left[\frac{\Gamma^+_{11}(\varepsilon_1)}{\Gamma^-_{11}(\varepsilon_1)}-\frac{\Gamma^+_{12}(\varepsilon_1)}{\Gamma^-_{12}(\varepsilon_1)}\right]
+\Gamma^-_{12}(\varepsilon_2)\left[\frac{\Gamma^+_{22}(\varepsilon_2)}{\Gamma^-_{22}(\varepsilon_2)}-\frac{\Gamma^+_{12}(\varepsilon_2)}{\Gamma^-_{12}(\varepsilon_2)}\right]=0,
\end{eqnarray}
which recovers the previous result in Ref.~\cite{swli2015aop}.

Moreover, it is known that the general solution at steady state is quite difficult.
Hence, we try to obtain analytical results in limiting regimes.
First, we gain the steady state populations at resonance ($\varepsilon_1=\varepsilon_2=\varepsilon$) within the two-terminal setup ($\gamma_M=0$), shown as
\begin{eqnarray}~\label{suprho12}
\rho^{ss}_{11}&=&\frac{1}{(\Gamma^-_{11}+\Gamma^-_{22})\mathcal{A}}
[(\Gamma^-_{11}+\Gamma^-_{22})\Gamma^+_{11}\Gamma^-_{22}+(\Gamma^+_{22}-\Gamma^+_{11})(\Gamma^-_{12})^2
-2\Gamma^-_{22}\Gamma^-_{12}\Gamma^+_{12}],\\
\rho^{ss}_{22}&=&\frac{1}{(\Gamma^-_{11}+\Gamma^-_{22})\mathcal{A}}
[(\Gamma^-_{11}+\Gamma^-_{22})\Gamma^-_{11}\Gamma^+_{22}+(\Gamma^+_{11}-\Gamma^+_{22})(\Gamma^-_{12})^2
-2\Gamma^-_{11}\Gamma^-_{12}\Gamma^+_{12}],\nonumber\\
\rho^{ss}_{gg}&=&[\Gamma^-_{11}\Gamma^-_{22}-(\Gamma^-_{12})^2]/\mathcal{A},\nonumber\\
\rho^{ss}_{12}&=&\frac{\Gamma^-_{11}\Gamma^-_{22}\Gamma^-_{12}}{(\Gamma^-_{11}+\Gamma^-_{22})\mathcal{A}}
(2\frac{\Gamma^+_{12}}{\Gamma^-_{12}}-\frac{\Gamma^+_{11}}{\Gamma^-_{11}}-\frac{\Gamma^+_{22}}{\Gamma^-_{22}}),\nonumber
\end{eqnarray}
with
$\Gamma^{\pm}_{ij}=\Gamma^{\pm}_{ij}(\varepsilon)$ and $\mathcal{A}=\Gamma^-_{11}(\Gamma^-_{22}+\Gamma^+_{22})+\Gamma^+_{11}\Gamma^-_{22}-\Gamma^-_{12}(\Gamma^-_{12}+2\Gamma^+_{12})$.
As the noise-induced coherence disappears $\Gamma^{\pm}_{12}(\varepsilon_i)=0~(i=1,2)$, the steady state solution is reduced to
\begin{eqnarray}
\rho^{ss}_{11}&=&\frac{\Gamma^+_{11}(\varepsilon_1)\Gamma^-_{22}(\varepsilon_2)}
{\Gamma^-_{11}(\varepsilon_1)(\Gamma^-_{22}(\varepsilon_2)+\Gamma^+_{22}(\varepsilon_2))+\Gamma^+_{11}(\varepsilon_1)\Gamma^-_{22}(\varepsilon_2)},\\
\rho^{ss}_{22}&=&\frac{\Gamma^-_{11}(\varepsilon_1)\Gamma^+_{22}(\varepsilon_2)}
{\Gamma^-_{11}(\varepsilon_1)(\Gamma^-_{22}(\varepsilon_2)+\Gamma^+_{22}(\varepsilon_2))+\Gamma^+_{11}(\varepsilon_1)\Gamma^-_{22}(\varepsilon_2)},\nonumber\\
\rho^{ss}_{gg}&=&\frac{\Gamma^-_{11}(\varepsilon_1)\Gamma^-_{22}(\varepsilon_2)}
{\Gamma^-_{11}(\varepsilon_1)(\Gamma^-_{22}(\varepsilon_2)+\Gamma^+_{22}(\varepsilon_2))+\Gamma^+_{11}(\varepsilon_1)\Gamma^-_{22}(\varepsilon_2)}.\nonumber
\end{eqnarray}

Next, within the three-terminal setup and in absence of the  noise-induced coherence ($\gamma_{12}(\varepsilon_i)=0$), we obtain the steady state populations  as
\begin{eqnarray}~\label{3tpop}
\rho^{ss}_{11}&=&[(\Gamma^-_{22}(\varepsilon_2)+\Gamma^+_M(\Delta))\Gamma^+_{11}(\varepsilon_1)+\Gamma^+_M(\Delta)\Gamma^+_{22}(\varepsilon_2)]/\mathcal{B},\\
\rho^{ss}_{22}&=&[(\Gamma^-_{11}(\varepsilon_1)+\Gamma^-_M(\Delta))\Gamma^+_{22}(\varepsilon_2)+\Gamma^-_M(\Delta)\Gamma^+_{11}(\varepsilon_1)]/\mathcal{B},\nonumber\\
\rho^{ss}_{gg}&=&[\Gamma^-_{22}(\varepsilon_2)\Gamma^-_{11}(\varepsilon_1)+\Gamma^-_{22}(\varepsilon_2)\Gamma^-_{M}(\Delta)
+\Gamma^+_{M}(\Delta)\Gamma^-_{11}(\varepsilon_2)]/B,\nonumber
\end{eqnarray}
with the coefficient
$\mathcal{B}=(\Gamma^+_{22}(\varepsilon_2)+\Gamma^-_{22}(\varepsilon_2)+\Gamma^+_{M}(\Delta))
(\Gamma^+_{11}(\varepsilon_1)+\Gamma^-_{11}(\varepsilon_1)+\Gamma^-_{M}(\Delta))-(\Gamma^-_M(\Delta)-\Gamma^+_{22}(\varepsilon_2))
(\Gamma^+_M(\Delta)-\Gamma^+_{11}(\varepsilon_1))$.

\section{Full counting statistics of the nonequilibrium V-type system}
To count the energy  flow into the bath $u$ at the time $\tau$, which starts from the $0$, the transferred heat is expressed as
${\Delta}q^u_\tau=\sum_k\omega_k{\Delta}n_{k,u}(\tau)$, with $\omega_k$ the phonon frequency in momentum $k$,
${\Delta}n_{k,u}=n_{k,u}(\tau)-n_{k,u}(0)$ and $n_{k,u}(t)$ the occupation phonon number in the bath $u$ at time $t$.
Then, we introduce the two-time measurement to analyze the currents.
Specifically, we include the measuring operator
$\hat{P}_{q^u_0}=|q^u_0{\rangle}{\langle}q^u_0|$ to detect the initial energy quantity
of the Hamiltonian $\hat{H}_u$ to be ${q}^u_0=\sum_k\omega_kn_{k,u}(0)$.
Similarly, at time $\tau$, we again measure $\hat{H}_u$ with the operator $\hat{P}_{q^u_\tau}=|q^u_\tau{\rangle}{\langle}q^u_\tau|$,
resulting in ${q}^u_\tau=\sum_k\omega_kn_{k,u}(\tau)$.
Hence, the joint probability for this two-time measurement is given by
\begin{eqnarray}~\label{subjp1}
\textrm{Pr}[{q}^u_\tau,{q}^u_0]=\textrm{Tr}
\{\hat{P}_{{q}^u_\tau}e^{-i\hat{H}\tau}\hat{P}_{{q}^u_0}\hat{\rho}_0\hat{P}_{{q}^u_0}e^{i\hat{H}\tau}\hat{P}_{{q}^u_\tau}\},
\end{eqnarray}
where $\hat{\rho}_0$ and $\hat{H}$ are the initial density matrix and Hamiltonian of the whole system, respectively.
By using the joint probability at Eq.~(\ref{jp1}), we define the probability of the transferred energy quant
${\Delta}Q^u_\tau$ during a finite time interval $\tau$ as
\begin{eqnarray}
\textrm{Pr}({\Delta}Q^u_\tau)=\sum_{{q}^u_\tau,{q}^u_0}
\delta[{\Delta}Q^u_\tau-(q^u_\tau-q^u_0)]\textrm{Pr}[{q}_u(\tau),{q}_u(0)].
\end{eqnarray}
Then, the generating function of the current statistics can be obtained as~\cite{mesposito2009rmp,mcampisi2011rmp}
\begin{eqnarray}
Z(\chi_u,t)&=&\int{d}{\Delta}Q^u_t\textrm{Pr}({\Delta}Q^u_t)e^{i\chi_u{\Delta}Q^u_t}\nonumber\\
&=&\textrm{Tr}\{e^{i\chi_u\hat{H}_u(0)}e^{-i\chi_u\hat{H}_u(t)}\hat{\rho}(0)\},
\end{eqnarray}
where $\chi_u$ is the counting field to count the flow into the bath $u$,
$\hat{H}_u(t)=\hat{U}^{\dag}(t)\hat{H}_u\hat{U}(t)$ and the evolution operator $\hat{U}(t)=e^{-i\hat{H}t}$.
Actually, the generating function can be alternatively expressed as~\cite{hmfriedman2018arxiv}
\begin{eqnarray}
Z(\chi_u,t)=\textrm{Tr}\{\hat{U}_{-\chi_u}(t)\hat{\rho}(0)\hat{U}^{\dag}_{\chi_u}(t)\}=\textrm{Tr}\{\hat{\rho}_{\chi_u}(t)\},
\end{eqnarray}
with $\hat{U}_{-\chi_u}(t)=e^{-i\chi_u\hat{H}_u/2}\hat{U}e^{i\chi_u\hat{H}_u/2}$
and $\hat{U}^{\dag}_{\chi_u}(t)=e^{i\chi_u\hat{H}_u/2}\hat{U}^{\dag}e^{-i\chi_u\hat{H}_u/2}$.
Hence, the cumulant generating function at steady state is
$G(\chi_u)=\lim_{t\rightarrow{\infty}}\frac{1}{t}Z(\chi_u,t)$.
The current fluctuations at steady state are obtained as
\begin{eqnarray}
J^{(n)}_u=\frac{{\partial^n}}{\partial(i\chi_u)^n}G(\chi_u)|_{\chi_u=0}.
\end{eqnarray}
Particularly, the steady state heat flux is the first cumulant
$J_u=\frac{{\partial}}{\partial(i\chi_u)}G(\chi_u)|_{\chi_u=0}$,
and the  noise power is the second cumulant
$S_{uu}=\frac{{\partial}^2}{\partial(i\chi_u)^2}G(\chi_u)|_{\chi_u=0}$.

For the nonequilibrium V-type system at Eq.~(\ref{hs0}), we count the energy current into the bath $u$, by adding a counting parameter set to
$\hat{H}$ as
$\hat{H}(\{\chi\})=e^{i\sum_{u}\chi_u\hat{H}_u/2}\hat{H}e^{-i\sum_{u}\chi_u\hat{H}_u/2}
=\hat{H}_s+\hat{H}_b+\hat{V}_M+\sum_{u=L,R}\hat{V}_u({\chi_u})$, where the modified system-bath interaction is given by
Eq.~(\ref{vu1}).
Based on the Born-Markov approximation, we obtain the quantum master equation at Eq.~(\ref{re1}).
Defining the vector expression of the density matrix $|\mathcal{P}(\{\chi\}){\rangle}=[\rho^{\{\chi\}}_{11},\rho^{\{\chi\}}_{22},\rho^{\{\chi\}}_{gg},\rho^{\{\chi\}}_{12},\rho^{\{\chi\}}_{21}]^\textrm{T}$,
the dynamical equation can be reexpressed in the Liouvillian framework as
\begin{eqnarray}
\frac{d}{dt}|\mathcal{P}(\{\chi\}){\rangle}=\mathcal{\hat{L}}(\{\chi\})|\mathcal{P}(\{\chi\}){\rangle}.
\end{eqnarray}
At steady state, the generating cumulant function is simplified as
\begin{eqnarray}
G(\{\chi\})=E_0(\{\chi\}),
\end{eqnarray}
where $E_0(\{\chi\})$ is the eigenvalue of the superoperator
$\mathcal{\hat{L}}(\{\chi\})$ with the maximal real part.
Hence, the heat current can be obtained as
\begin{eqnarray}~\label{supju1}
J_u=\frac{{\partial}E_0(\{\chi\})}{{\partial}(i\chi_u)}|_{\{\chi\}=0}
={\langle}I|\frac{{\partial}\mathcal{\hat{L}}(\{\chi\})}{{\partial}(i\chi_u)}|_{\{\chi\}=0}|\mathcal{P}_{ss}{\rangle},
\end{eqnarray}
where ${\langle}I|=[1,1,1,0,0]$ is the unit vector and
$|\mathcal{P}_{ss}{\rangle}=[\rho^{ss}_{11},\rho^{ss}_{22},\rho^{ss}_{gg},\rho^{ss}_{12},\rho^{ss}_{21}]^\textrm{T}$ is the steady state of V-type system.
Finally, the heat currents are obtained at Eq.~(10-12).

\section{Steady state particle currents}
By applying a similar scheme to count the particle flow, the transformed Hamiltonian is given by
$\hat{H}(\{\chi_p\})=e^{i\sum_{u}\chi^p_u\hat{N}_u/2}\hat{H}e^{-i\sum_{u}\chi^p_u\hat{N}_u/2}
=\hat{H}_s+\hat{H}_b+\hat{V}_M+\sum_{u=L,R}\hat{V}_u({\chi^p_u})$, where $\hat{N}_u=\sum_k\hat{a}^{\dag}_{k,u}\hat{a}_{k,u}$ and the modified system-bath interaction is given by
\begin{eqnarray}
\hat{V}_u(\chi^p_u)=\sum_{k,i}(g^i_{k,v}e^{i\chi^p_u}\hat{a}^{\dag}_{k,u}|g{\rangle}{\langle}e_i|
+g^{i*}_{k,u}e^{-i\chi^p_u}\hat{a}_{k,u}|e_i{\rangle}{\langle}g|).
\end{eqnarray}
Based on the second-order perturbation, the modified master equation is given by
\begin{eqnarray}~\label{supre2}
\frac{d\hat{\rho}_{\{\chi_p\}}}{dt}&=&-i[\hat{H}_s,\hat{\rho}_{\{\chi_p\}}]\\
&&-\frac{1}{2}\sum_{i,j;\sigma=\pm}\Gamma^{\sigma}_{ij}(\varepsilon_j)(\hat{\phi}^{\overline{\sigma}}_i\hat{\phi}^{{\sigma}}_j\hat{\rho}_{\{\chi_p\}}
+\hat{\rho}_{\{\chi_p\}}\hat{\phi}^{\overline{\sigma}}_j\hat{\phi}^{{\sigma}}_i)\nonumber\\
&&+\frac{1}{2}\sum_{i,j;\sigma=\pm}(\Gamma^{\sigma}_{ij}(\varepsilon_i,\{\chi_p\})+\Gamma^{\sigma}_{ij}(\varepsilon_j,\{\chi_p\}))
\hat{\phi}^\sigma_i\hat{\rho}_{\{\chi_p\}}\hat{\phi}^{\overline{\sigma}}_j\nonumber\\
&&+\frac{1}{2}\sum_{\sigma=\pm}\Gamma^{\sigma}_m(\Delta)
([\hat{\psi}^\sigma\hat{\rho}_{\{\chi_p\}},\hat{\psi}^{\overline{\sigma}}]+[\hat{\psi}^{{\sigma}},\hat{\rho}_{\{\chi_p\}}\hat{\psi}^{\overline{\sigma}}])\nonumber
\end{eqnarray}
where the modified transition rates are
$\Gamma^+_{ij}(\omega,\{\chi_p\})=\sum_v\gamma^v_{ij}n_v(\omega)e^{-i\chi_v}$ and
$\Gamma^-_{ij}(\omega,\{\chi_p\})=\sum_v\gamma^v_{ij}(1+n_v(\omega))e^{i\chi_v}$.
Then, the particle currents can be obtained as
\begin{eqnarray}~\label{jp1}
J^p_L&=&\sum_{j=1,2}\gamma^L_{jj}[(1+n_L(\varepsilon_j))\rho^{ss}_{jj}-n_L(\varepsilon_j)\rho^{ss}_{gg}]\\
&&+\frac{1}{2}[\sum_{j=1,2}\gamma^L_{12}(1+n_L(\varepsilon_j))](\rho^{ss}_{12}+\rho^{ss}_{21}),\nonumber\\
J^p_R&=&\sum_{j=1,2}\gamma^R_{jj}[(1+n_R(\varepsilon_j))\rho^{ss}_{jj}-n_R(\varepsilon_j)\rho^{ss}_{gg}]\nonumber\\
&&+\frac{1}{2}[\sum_{j=1,2}\gamma^R_{12}(1+n_R(\varepsilon_j))](\rho^{ss}_{12}+\rho^{ss}_{21}),\nonumber\\
J^p_M&=&\gamma_M(1+n_M(\Delta))\rho^{ss}_{11}-\gamma_Mn_M(\Delta)\rho^{ss}_{22}.\nonumber
\end{eqnarray}
The particle conservation law results in $J^p_L+J^p_R=0$, which can be obtained from Eq.~(\ref{supre0}) by setting $\frac{d}{dt}\rho_{ij}=0$.

\section{A systematical perturbation method to obtain steady state current culumants}
The cumulant generating function is $G(\chi)=E_{\chi}$, where $E_\chi$ is the eigenvalue with maximal real part
$\hat{H}_{\chi}|P_{\chi}{\rangle}=E_\chi|P_{\chi}{\rangle}$.
Then, we expand all terms as
$\hat{H}_{\chi}=\sum^{\infty}_{n=0}\frac{(i\chi)^n}{n!}\hat{H}_n$,
${E}_{\chi}=\sum^{\infty}_{n=0}\frac{(i\chi)^n}{n!}{E}_n$,
and
$|P_{\chi}{\rangle}=\sum^{\infty}_{n=0}\frac{(i\chi)^n}{n!}|P_{n}{\rangle}$.
If we consider the order $(i\chi)^N$, we obtain
\begin{eqnarray}
\sum^N_{n=0}\frac{\hat{H}_n}{n!(N-n)!}|P_{N-n}{\rangle}=\sum^N_{k=0}\frac{E_k}{k!(N-k)!}|P_{N-k}{\rangle},
\end{eqnarray}

For the zeroth order, the steady state is obtained as
\begin{eqnarray}
\hat{H}_0|P_0{\rangle}=0,\nonumber
\end{eqnarray}
and the left eigenvector is given by ${\langle}I|\hat{H}_0=0$.
Consequently, the $N$th order cumulants is given by
\begin{eqnarray}
E_N&=&\sum^N_{n=1}\frac{N!}{n!(N-n)!}{\langle}I|\hat{H}_n|P_{N-n}{\rangle}\nonumber\\
&&-\sum^{N-1}_{k=1}\frac{N!}{k!(N-k)!}E_k{\langle}I|P_{N-k}{\rangle},
\end{eqnarray}
and the corresponding state is
\begin{eqnarray}
|P_N{\rangle}=\hat{R}\sum^N_{n=1}\frac{N!}{n!(N-n)!}(E_n-\hat{H}_n)|P_{N-n}{\rangle},
\end{eqnarray}
with $\hat{R}=\hat{Q}\hat{H}^{-1}_0\hat{Q}$, $\hat{H}^{-1}_0$ the Moore-Penrose inverse, and $\hat{Q}=1-|P_0{\rangle}{\langle}I|$ to eliminate the singular value of $\hat{H}_0$.

Specifically,
\begin{eqnarray}
E_1&=&{\langle}I|\hat{H}_1|P_0{\rangle}\\
|P_1{\rangle}&=&\hat{R}(E_1-\hat{H}_1)|P_0{\rangle},\nonumber\\
E_2&=&2{\langle}I|(\hat{H}_1-E_1)|P_1{\rangle}+{\langle}I|\hat{H}_2|P_0{\rangle}\nonumber\\
&=&{\langle}I|\hat{H}_2|P_0{\rangle}-2{\langle}I|(\hat{H}_1-E_1)\hat{R}(\hat{H}_1-E_1)|P_0{\rangle}\nonumber\\
|P_2{\rangle}&=&2\hat{R}(E_1-\hat{H}_1)|P_1{\rangle}+\hat{R}(E_2-\hat{H}_2)|P_0{\rangle}\nonumber
\end{eqnarray}
Hence, the steady state flux is given by $J=E_1$ and the noise is $S=E_2$.

\end{document}